\newcommand{\er}{\hat {\bf e}_{r}}

\newcommand{\eph}{\hat {\bf e}_{\phi}}
\newcommand{\ez}{\hat {\bf e}_{z}}

\newcommand{\deph}{\dot{\hat{\bf e}}_{\phi}}

\documentclass[12pt,prb,aps,preprint]{revtex4}

\begin{document}

\title{Tracing a planet's orbit with a straight hedge and a
compass with the help of the hodograph and the Hamilton vector}

\author{E. Guillaum\'{\i}n-Espa\~na}
\affiliation{Laboratorio de Sistemas Din\'amicos,
Departamento de Ciencias B\'asicas, Universidad
Aut\'onoma Metropolitana-Azcapotzalco, Apartado
Postal 21-267, Coyoac\'an 04000 D.\ F., M\'exico}

\author{H. N. N\'u\~nez-Y\'epez}
\affiliation{Departamento de F\'{\i}sica,
Universidad Aut\'onoma Metropolitana-Iztapalapa,
Apar\-tado Postal 55-534, Iztapalapa 09340 D.\ F., M\'exico}

\author{A. L. Salas-Brito}
\affiliation{Laboratorio de Sistemas Din\'amicos, Departamento de
Ciencias B\'asicas, Universidad Aut\'onoma
Metropolitana-Azcapotzalco, Apartado Postal 21-267, Coyoac\'an
04000 D.\ F., M\'exico}
\email{asb@correo.azc.uam.mx}

%\date{\today}

\begin{abstract}
We describe a geometrical method for tracing a planet's orbit 
using its velocity hodograph, that is, the path of the planet's 
velocity. The method requires only a straight edge, a compass, 
and the help of the hodograph. We also obtain analitically the 
hodograph and some  of the features of the motion that can be 
obtained from it. \end{abstract}

\maketitle

%\pacs{}

\section{Introduction}
One way of solving the Kepler problem is to determine the path of
the planet in velocity space. This path is called the
hodograph.\cite{hamilton} Although this way of solving the problem
apparently involves a detour, it ends up being one of the simplest
ways of finding the
orbit.\cite{milnor,gutz90,moreno,abelson,us96,butikov00,
derbes01,maxwell,
thomsontait,anicin,feynman,sivardiere,us98,newtonianrmf,
scatteringrmf,ejp00}.  Moreover,
this approach makes it straightforward to obtain an additional
constant of the motion, namely the Hamilton vector. A recent
contribution to using the hodograph is the interesting paper by
Derbes.\cite{derbes01} This discussion is beautifully conducted
using Euclidian geometry, but some people may be confounded by the
use of such arguments. 

We discuss a different geometrical approach for tracing a planet's
orbit that also uses the hodograph. Our method can  be
cast as a series of steps starting from the initial position and
velocity of the planet. Then using the Hamilton vector and the
hodograph, the method directly leads to the points on the planetary
orbit.  In addition to constructing the orbits, we  discuss
other features of the motion that follow naturally from our
approach.

The paper is organized as follows. In Sec.~II the analytical basis
of the hodographic method is expounded. We introduce the Hamilton
vector, an interesting conserved quantity that is
rarely used today. We then obtain the hodograph and some of the
features of the motion that can be obtained from it. The geometrical
method for determining the orbit is the subject of Sec.~III. We
describe a straight edge and compass method for finding both the
hodograph and the planet's orbit starting from its initial position
and velocity. The technique can be made simple enough to
be used to teach some of the underpinnings of planetary motion
 --- especially if it is used with
a computer program for visualizing the
constructions.\cite{butikov99,geometers} 
 In
Sec.~IV we show geometrically that the orbit is indeed elliptical
and that the direction of the velocities calculated from it
correspond to those directly obtained from the hodograph. Finally,
in Sec.~V we make some concluding remarks and  suggest some
problems that can be solved with the methods presented in the paper.

\section{The analytic approach to the hodograph}\label{sec2}
The geometrical arguments\cite{moreno,derbes01} used for establishing
the circularity of the Kepler problem hodograph can be stated
analytically as follows.\cite{us96,ejp00} The equation of motion of a
planet of mass $m$ in the gravitational field of the sun is
\begin{equation}\label{1}
m \frac {d{\bf v}} {dt}= - \,\frac {GMm} {r^2} \er
\end{equation}
where $M$ is the mass of the sun, $r$ is the distance between the
sun and the planet, and $\er$ is a unit vector that points in the
radial direction from the sun toward the planet. We work in a
polar coordinate system where the basis vectors $\er$, $\eph$, and
$\ez$ satisfy $\er =\eph \times \ez$. Hence, using the polar
identity $\deph=-
\dot\phi \er$ and the conservation of angular momentum
(so that $L=mr^2\dot\phi$ is a constant), we can eliminate $r$ and
the unit vector
$\er$, and replace them by $\phi$ and $\deph$. In this way we can
express Eq.~(\ref{1}) in the form
\begin{equation}
\label{2}
\frac{d}{dt}({\bf v}-u \eph)=\frac{d}{dt}({\bf v}-{\bf
u})=0,
\end{equation}
where we have defined the constant $u \equiv GMm/L$ which has
dimensions of velocity and the rotating vector of constant magnitude
${\bf u} = u \eph$.\cite{ejp00,butikov99} Equation~(\ref{2})
shows that the Hamilton vector,
\begin{equation}\label{h}
{\bf h}\equiv {\bf v}-{\bf u} ={\bf v} + {u\over r L} ({\bf r} \times
{\bf L}),
\end{equation}
is a constant of the motion in the Kepler problem.

The constancy of ${\bf h}$ is an exclusive feature of the
$1/r^2$ force  and is related to its extraordinary
symmetries.\cite{us92} Equation~(\ref{2}) shows that ${\bf h}$ is
in the plane of the orbit and orthogonal to ${\bf L}$.
Equation~(\ref{h}) also shows that the hodograph has a dynamical
symmetry axis that is defined by ${\bf h}$. The axis is dynamical
in the sense that the constant Hamilton vector (\ref{h}) is defined
using the dynamical variables ${\bf r}$, and ${\bf v}$. Hence, the
orbit also has to have a dynamical symmetry axis which must be
orthogonal to
${\bf h}$. The orbit's symmetry axis can be defined by any vector
constant orthogonal to ${\bf h}$ and in the orbital plane. An
obvious choice is
${\bf A} = {\bf h}\times{\bf L}={\bf v}\times {\bf L}-{\bf r}(uL/r)
$, which is the Laplace or Runge-Lenz vector.\cite{us92} Another
consequence of Eq.~(\ref{h}) is that every bounded orbit has to be
not only planar but periodic.

{}From the definition (\ref{h}) we can express the velocity as a
rotating vector with fixed magnitude, $u\,\eph$, superimposed on the
constant Hamilton vector
\begin{equation}\label{hodo}
{\bf v}= {\bf h} + u \eph.
\end{equation}
As can be seen from Eq.~(\ref{hodo}) (by evaluating {\bf h} for
$\phi=0$),
${\bf h}$ is parallel to the velocity at the perihelion ${\bf v}_p$.
Equation~(\ref{hodo}) also shows that the
hodograph is an arc
of a circle of radius
$u$ with its center located at the tip of ${\bf h}$, that is, it
is not centered at the origin in velocity space. The hodograph is
always concave toward the origin when the interaction is attractive
as is the case of planetary motion (see Fig.~1).

The velocity vector does not necessarily traverse the entire
hodographic circle during the motion; it may just move on a
circular arc. To see this, we write Eq.~(\ref{hodo}) in cartesian
coordinates using ${\bf h}$ to define the
$x$-axis:
$v_x=u
\sin
\phi$ and $v_y=u \cos \phi + {h}$, where $\phi$ is the angle
between
${\bf u}$ and ${\bf h}$. These expressions for $v_x$ and $v_y$
imply that the equation of the hodograph can be written as
$v_x^2 + (v_y-h)^2 =u^2$, which represents a circle of radius $u$
centered at the point $(0, h)$ in velocity space. If we make a
simple substitution for the components, the speed
$v=\sqrt{v_x^2+v_y^2}$ can be written as
\begin{equation}
\label{speed}
v=u\sqrt{1+\epsilon^2-2\epsilon \cos \phi},
\end{equation}
where $\epsilon\equiv h/u$. Equation~(\ref{speed}) shows that the
polar angle is limited to the interval $|\phi| \le
\phi_{\rm max}$ when $\epsilon \ge 1$, where $\phi_{\rm max}
\equiv\arccos(1/\epsilon)$. That is, the hodograph coincides with the
entire circle only when $\epsilon < 1$. When
$|\phi_{\rm max}|$ is approached, the
velocity becomes tangent to the hodograph and the speed reaches a
limiting value
$v_\infty=u\sqrt{\epsilon^2-1}$. We  conclude from this
expression  that when
$\epsilon > 1$, the planet is unbounded and moves asymptotically (as
$t\to \infty$) toward a point in velocity space that we
correspondingly call ${\bf v}_\infty$.

Another interesting relation exists between the hodograph radius
$u$ and the speeds at the perihelion $v_p$ and at the aphelion
$v_a$:
\begin{equation}
\label{u}
u=\frac {1} {2} (v_a + v_p).
\end{equation}
To derive Eq.~(\ref{u}) we need to equate the total energy evaluated at
these two special positions on the orbit, solve the
resulting equation for $u$, and then substitute the angular momentum
evaluated at the perihelion. Moreover, ${\bf h}$ can be expressed in
terms of the sum of these two velocities: ${\bf h}= ({\bf v}_p +
{\bf v}_a)/2$, and its magnitude can be expressed in terms of the
difference of the corresponding speeds:
$h=(v_p-v_a)/2$.\cite{tiberiis} These results are easily seen
geometrically from Fig.~1.

The orbit of the planet can be obtained by projecting ${\bf h}$ onto
$\eph$ to obtain
\begin{equation}
\label{orbit}
r=\frac {L/m} {u+ h\, \cos \phi},
\end{equation}
which is the polar equation of a conic with semilatus rectum $L/mu$
and eccentricity $\epsilon=h/u$. The angle $\phi$, which has the same
meaning in both Eqs.~(\ref{orbit}) and (\ref{speed}), is usually
called the true anomaly in celestial mechanics. Therefore, the
possible orbits are ellipses when $h >u$ ($\epsilon<1)$,
parabolas when $h=u$ ($\epsilon=1)$, and hyperbolas when
$h<u$ ($\epsilon>1)$. If the orbit is elliptical, the
hodograph traverses the entire circle. In any other case the
hodograph traverses just an arc of the circle --- although in the
parabolic case, it only misses a single point on it.

By using Eqs.~(\ref{speed}) and (\ref{orbit}) in the bounded case, we
can easily check that $v_p/v_a=r_a/r_p$, where $r_a$ and $r_p$ are,
respectively, the distances to the planet at the aphelion and
perihelion. This relation also follows from angular momentum
conservation. We may also write the energy of the planet as
\begin{equation}\label{E}
E=\frac {m} {2} (h^2-u^2).
\end{equation}
{}From Eq.~(\ref{E}) we immediately see that the orbit is elliptical
if $E<0$, parabolic if $E=0$, and hyperbolic if $E>0$.
We may also see from Eq.~(\ref{E}) that in the hyperbolic case, we can
write $h^2=u^2+v_\infty^2$. That is, the limiting velocity ${\bf
v}_\infty$, the Hamilton vector ${\bf h}$, and the limiting vector
${\bf u}$ as $t\to \infty$ (${\bf u}_\infty$), always form a right
triangle with ${\bf h}$ as the hypotenuse. This result comes in handy
for deriving geometrically the Rutherford relation for the
scattering of celestial bodies off the
sun,\cite{newtonianrmf,scatteringrmf} as is the case for comets
moving in hyperbolic orbits.

\section{Tracing the orbit from the initial conditions}\label{sec3}
Geometrical methods are powerful and intuitive,\cite{chandra} although
some students may find them unfamiliar and hence confusing.
Nevertheless, these methods can be used to find ${\bf h}$, the
hodograph, and then to trace the orbit starting with an initial
position
${\bf r}_0$ and velocity ${\bf v}_0$. If this method is properly
presented, it can be very concrete because students can draw, point
by point, any orbit by themselves.

The method can be described as follows: Given ${\bf r}_0$ and
${\bf v}_0$, we can obtain the magnitude of the planet's angular
momentum $L=mr_0 v_0 \sin \delta$, where $\delta$ is the angle between
${\bf r}_0$ and ${\bf v}_0$ (see Fig.~1), or as the area spanned by
these same two vectors (area $VOO'C$ in Fig.~1). Once $L$ is known,
the hodograph radius $u=GMm/L$ can be calculated.

We next select a point $F$ on the plane as the origin of the
coordinates, that is, as the position of the center of force. From
this origin we draw a line segment $\overline{FR}$ (parallel to
${\bf r}_0$) representing the initial position. The line
$\overline{FR}$ can be extended to the position we choose for the
velocity space origin $O$. From $O$ we also draw the segment
$\overline{OV}$ corresponding to the initial velocity. Then we
draw, perpendicular to $\overline{FR}$, a line segment
$\overline{OO'}$ of length $u$. Using the parallelogram rule, we
add the segments $\overline{OV}$ and $\overline{OO'}$ to obtain the
line $\overline{OC}$ corresponding to Hamilton's vector. It is now
a matter of tracing a circle of radius $u$ centered at point $C$,
the tip of Hamilton's vector. This circle is the hodograph. Notice
that the points marked $V_p$ and $V_a$ correspond, respectively, to
the velocities at the perihelion and at the aphelion. The velocity
vectors at the aphelion and perihelion are necessarily orthogonal to
the symmetry axis ($\overline{FP}$) of the orbit. This
construction is illustrated in Fig.~1. The symmetry axis has the
direction of the Runge-Lenz vector (shown in Fig.~2). 

Figures~2, 3, and 4 include the same information as Fig.~1, but
have certain features that have been added or removed to focus the
reader on a particular point. We have packed much information in
Figs.~1 and 2. Angular momentum conservation is explicitly included
because we have assumed that the orbit lies in a plane. The
flatness of all the orbits can be shown to imply the central nature
of the force.\cite{urbankte}

Given the amount of information in Fig.~1, it is not
surprising that we can determine from Fig.~1 the bounded or
unbounded nature of the orbit stemming from ${\bf r}_0$ and
${\bf v}_0$: if the point $O$, the origin in velocity space,
is within the hodographic circle, the orbit is necessarily elliptical
and hence bounded, otherwise, the orbit is hyperbolic or parabolic
(and hence unbounded). The parabolic case only occurs if $O$ sits
exactly on the hodographic circle, a property that follows directly from
Eq.~(\ref{E}). For a circular orbit the center of the hodograph
$C$ coincides with the velocity space origin $O$, that is, $h=0$, which
means that the speed equals the constant hodograph radius $u$.

To completely determine the orbit (elliptical, in this case) with
the information shown in Fig.~1, we proceed as follows (see
Fig.~2). Trace the line $\overline{FP}$ that is perpendicular to
the line
$\overline{OC}$ and passes through $F$. This line is the symmetry axis
of the orbit, as follows from the orthogonality property mentioned
earlier. By using the holograph and the symmetry axis of the orbit,
we can begin to locate points on the planet's orbit. In Fig.~2 the
points marked $O$,
$V$,
$F$, $R$, and $C$, have the same interpretation as in Fig. 1; for
example, the segment $\overline{OC}$ represents the Hamilton vector
${\bf h}$.

To locate any point on the orbit, extend the line $\overline{VO}$ back
until it intercepts the hodograph at point $V_s$. Trace a segment that
is perpendicular to $\overline{CV_s}$ and passes through $R$. This line
intercepts the symmetry axis at the point $F'$. To locate the point on
the orbit corresponding to any given point on the hodograph, we
notice that we already have one such pair of points, namely the initial
conditions (points $R$ and $V$ in Fig.~2). We choose another
point, $V'$, on the hodograph, and extend the straight line $\overline
{OV'}$ until it again intersects the hodograph at point
$V_s'$. Draw two perpendiculars, one to $\overline{CV'}$ and the other
to $\overline{CV_s'}$, passing through $F$ and $F'$, respectively. The
intersection, $R'$, of these two perpendiculars is the required point
on the orbit. This construction is similar to the case of
lines
$\overline{CV}$ and
$\overline{CV_s}$ that meet at the initial condition $R$.

This process is repeated for every point on the hodograph and in
this way we can trace the complete orbit starting with the initial
conditions and using only a straight edge and a compass. This last
feature is a manifestation of the extreme regularity of the
orbit.\cite{us92} We should note that parabolic and hyperbolic
orbits can also be traced using variants of the method described
above.

\section{The shape of the orbit and the associated
velocities}\label{sec4}
Our method for drawing an orbit is fully contained in Sec.~III.
Here we will address two loose ends that are not important if you
are only interested in tracing the orbit. What is the shape of the
orbit and how can we be sure that the points we have found on the
orbit are the required points for the velocity vectors as required
by the dynamics?

That the loci of the points found by the method of 
Sec.~\ref{sec3} is indeed an ellipse can be seen as follows.
We first draw a circular arc $F'W$ centered at point $R$ with radius
$\overline{RF'}$; this arc helps trace an auxiliary circle
centered at
$F$ with a radius equal to the sum of the lengths of the lines
$\overline{FR}$ and $\overline{RF'}$, that is, the length of
$\overline{FW}$. This radius equals the length of the orbit's major
axis, that is, the line $\overline {P_aP_p}$, in Fig.~3. The points
$W$ and
$W'$ are, respectively, the intersections of lines $\overline{FR}$ (the
initial condition) and
$\overline{FR'}$ (the calculated point on the orbit) with the auxiliary
circle.

The isosceles triangles $\triangle CV'V_s'$
and
$\triangle R'W'F'$ in Fig.~3 are similar to each other, because the
line
$\overline{FR'}$ makes the same angle with line $\overline{FF'}$ as the
line $\overline{CV'}$ makes with the line $\overline{OC}$. Thus the
point
$R'$ on the orbit is at the same distance from both $W'$ and $F'$. So
we must have that $\overline{FR'} + \overline{R'F'} = \overline{FR} +
\overline{RF'}=\overline{P_aP_p}$. This description shows that the
sum of the distances from points on the orbit, such as $R$ and $R'$, to
the points
$F$ and $F'$ is a constant; this equality is precisely the defining
property of an ellipse. Therefore, the planets travel on elliptical
paths and the sun (the origin) is located at the position of one of
the foci of the ellipse, $F$ in Figs.~1, 2, and 3; $F'$ is the
other focus of the ellipse.

Once the shape of the orbit has been established, we can check
that the velocities, as defined by points on the hodograph, are
parallel to the tangents at the corresponding points on the ellipse.
For example, in Fig.~3, point $V$ on the hodograph and the tangent to
the ellipse at point $R$ are parallel.

Figure~4 is similar to Fig.~3 with
certain lines added and others removed with the purpose of explaining
what follows. The argument relies on identifying the three
similar triangles, $\triangle WFW'$, $\triangle WRF'$, and
$\triangle F'R_sW_s$, which by construction are isosceles, and on the fact that a triangle
inscribed in a circle (for example, $\triangle WW'W_s$) whose
diameter coincides with one of the sides of the triangle, is
necessarily a right triangle. Trace the lines $\overline{WW_s}$
through $F'$,
$\overline{W_sW''}$ through $F$, and $\overline{W''W}$ which
closes the right triangle. Now trace lines $\overline{MR}$ and
$\overline{M_sR_s}$. These are, respectively, perpendicular
bisectors of the lines $\overline{WF'}$ and
$\overline{F'W_s}$, and at the same time, bisectors of the angles
$\angle WRF'$ and $\angle F'R_sW_s$. These properties guarantee
that $\overline {MR}$ and $\overline{M_sR_s}$ are parallel to the
the tangents to the orbit at the points $R$ and $R_s$. To see that
$\overline{MR}$ and
$\overline{M_sR_s}$ are parallel to $\overline{VO}$ and
$\overline{OV_s}$, respectively, it suffices to understand that
$\overline{WF'}$ is perpendicular to $\overline{VV_s}$.
This argument establishes that the tangent to any point on the
orbit is necessarily parallel to the corresponding velocity on the
hodograph.

\section {Concluding Remarks}
The geometrical approach described in this paper is simple and
direct and can serve to explain, even to beginning students, how to
trace a planet's orbit from the initial conditions. If presented
with no further explanations, the approach may be regarded as being
similar to the  idea described by the Mayan astronomer analogy
as told by Richard Feynman.\cite{feynman85} Our method is an
attempt to exhibit in simple terms the geometric beauty of
dynamics --- beauty that captured the heart of Newton
himself.\cite{milnor,moreno,chandra,us98}

Our  method can be profitably applied to
other related problems. For example, what if we wish to describe
the trajectory of a comet? How can the hodographic method be used
when the initial conditions lead to a hodographic circle that
does not surround the velocity space origin $O$? The extension of
the method to parabolic or hyperbolic orbits can be a
relatively simple project for interested students. Would it be
possible to account for the untraversed branch of the hyperbola?
Can this branch have some physical interpretation?

We also can  take advantage of the right triangle formed by ${\bf
h}$,
${\bf u}$, and ${\bf v}_\infty$ to derive the cross section for
comets bouncing off the sun. To begin with, 
the angle
$\xi$ between the vectors ${\bf v}_{+\infty}$ and ${\bf
v}_{-\infty}$ is the scattering angle. Then, after drawing the
hodograph, the Hamilton vector ${\bf h}$, and the two velocity
vectors ${\bf v}_{\pm\infty}$, it is a simple matter of geometry 
to obtain the relation
\begin{equation}
\label{rutherford}
L=\frac {GM} {v_\infty} \cot(\xi/2),
\end{equation}
from which the scattering cross section follows.\cite{ll}
Equation~(\ref{rutherford}) is usually called the Rutherford
relation.\cite{scatteringrmf}

A slightly more ambitious project would be to obtain astronomical
data for a planet and use this information to determine the initial
position and the initial velocity of the planet. Then the student
could trace the orbit with the method described in this paper and
then to check  such theoretical orbit against the experimental
data. That is, to compare the orbit determined directly from the
astronomical data with the orbit traced with our method starting
from just the initial conditions. A direct way of performing this
comparison is to use widely distributed data as that in the Orbit
of Mars Experiment of the Project Physics Course \cite{ppc}, which
includes photography of the night sky containing planet Mars and
detailed instructions for reconstructing its orbit from such images.

\begin{acknowledgements}
HNN-Y and ALS-B have been partially supported by a PAPIIT-UNAM research
grant. We acknowledge with thanks the Ricardo J. Zevada Foundation
for the graphics software used in this work.
We also want to thank P. Weiss, M. Percy, E. Hera, J. E. Juno, R.
Zeus, K. Bielii, G. Gorbe, M. Botitas, Z. Ita, M. Crida, P.
Schwartz, A. Saltar\'{\i}n, C. Schr\"odie, and P. M. Zura and her family. We dedicate
this work to the memories of F. C. Bonito (1987--2002), M. Osita
(1990--2001), and Ch.\ Shat
(1991--2001).
\end{acknowledgements}

\newpage

\begin{figure}[h]
\caption{The circular hodograph and the Hamilton vector ${\bf h}$
for planetary motion. We exhibit the initial position ${\bf r}_0
$, the initial velocity ${\bf v}_0$, and the rotating vector with
constant magnitude ${\bf u}$. We also show the polar unit vectors
$\er$ and $\eph$, the vector $\ez$ points upward from the plane of
the paper.
$V_p$ and $V_a$ are the points on the hodograph corresponding to
the velocities at the perihelion and at aphelion, respectively. The
figure shows that $u=(v_a+v_p)/2$, ${\bf h}=({\bf v}_a+{\bf
v}_p)/2$, and $h=(v_a-v_p)/2$. Note that by looking at the figure
you can determine the shape of the planet's orbit. If $O$ is within
the hodographic circle, as in the case illustrated here,
$E<0$ and the orbit is elliptical. If $O$ sits on the hodographic
circle, then the energy vanishes $E=0$ and the orbit is parabolic.
And if $O$ is outside the circle, then $E>0$ and the orbit is
hyperbolic.}
\end{figure}

\begin {figure}[h]
\caption{The geometrical method for tracing the orbit. $F$ is the position of the center of force, $C$ is
the hodograph center,
$O$ is the origin in velocity space, and the line $\overline{FF'}$
is the orbit axis of symmetry, that is, the direction of the
Runge-Lenz vector
${\bf A}$. Note that every point on
the hodograph (for example, ${\bf v}$) corresponds to a point on
the orbit (${\bf r}$). The symbols have the same
meaning as the corresponding ones in Fig.~1. $P$ is {\sl not}
necessarily an apsidal point on the ellipse.}
\end{figure}

\begin{figure}[h]
\caption{We redraw Fig.~2 to show that the orbit is an ellipse.
The shaded triangles are both isosceles and similar to each other.
These properties are useful for showing that the sum of the
distances from any point on the orbit to the points $F$ and $F'$
is a constant.}
\end{figure}

\begin{figure}[h]
\caption{Illustration of the fact that the velocity, as taken from
the hodograph, is always parallel to the tangent at the
corresponding point on the orbit. For example, the velocity
corresponding to the line
$\overline{OV}$ is parallel to the tangent line $\overline{MR}$ at
the point $R$, and the to velocity, corresponding to the
line
$\overline{OV_s}$, is parallel to the line $\overline{M_sR_s}$
which is tangent to the orbit at point $R_s$. The three shaded
triangles are similar to each other. }
\end{figure}

\end{document}